\documentclass[traditabstract]{aa}
\usepackage[varg]{txfonts}
\usepackage{graphicx}
\usepackage{natbib}
\bibpunct{(}{)}{;}{a}{}{,}

\begin{document}

\title{Period and amplitude variations in post-common-envelope eclipsing binaries observed with SuperWASP\thanks{Tables~\ref{aadordat}--\ref{qsvirextra} are available in electronic form at the CDS via anonymous ftp to cdsarc.u-strasbg.fr (130.79.128.5) or via http://cdsweb.u-strasbg.fr/cgi-bin/qcat?J/A+A/}}

\author{M.~E.~Lohr\inst{\ref{inst1}}\and
  A.~J.~Norton\inst{\ref{inst1}}\and
  D.~R.~Anderson\inst{\ref{inst2}}\and
  A.~Collier~Cameron\inst{\ref{inst3}}\and
  F.~Faedi\inst{\ref{inst4}}\and C.~A.~Haswell\inst{\ref{inst1}}\and\\
  C.~Hellier\inst{\ref{inst2}}\and S.~T.~Hodgkin\inst{\ref{inst5}}\and
  K.~Horne\inst{\ref{inst3}}\and U.~C.~Kolb\inst{\ref{inst1}}\and
  P.~F.~L.~Maxted\inst{\ref{inst2}}\and\\
  D.~Pollacco\inst{\ref{inst4}}\and I.~Skillen\inst{\ref{inst6}}\and
  B.~Smalley\inst{\ref{inst2}}\and R.~G.~West\inst{\ref{inst4}}\and
  P.~J.~Wheatley\inst{\ref{inst4}}} \institute{Department of Physical
  Sciences, The Open University, Walton Hall, Milton Keynes MK7\,6AA,
  UK\\ \email{Marcus.Lohr@open.ac.uk}\label{inst1}\and Astrophysics
  Group, Keele University, Staffordshire ST5\,5BG, UK\label{inst2}\and
  SUPA, School of Physics \& Astronomy, University of St. Andrews,
  North Haugh, Fife KY16\,9SS, UK\label{inst3}\and Department of
  Physics, University of Warwick, Coventry CV4\,7AL,
  UK\label{inst4}\and Institute of Astronomy, Madingley Road,
  Cambridge CB3\,0HA, UK\label{inst5}\and Isaac Newton Group of
  Telescopes, Apartado de Correos 321, E-38700 Santa Cruz de la Palma,
  Tenerife, Spain\label{inst6}} \date{Received 17 April 2014/ Accepted 19 May 2014}

\abstract {Period or amplitude variations in eclipsing binaries may
  reveal the presence of additional massive bodies in the system, such
  as circumbinary planets.  Here, we have studied twelve
  previously-known eclipsing post-common-envelope binaries for
  evidence of such light curve variations, on the basis of multi-year
  observations in the SuperWASP archive.  The results for HW~Vir
  provided strong evidence for period changes consistent with those
  measured by previous studies, and help support a two-planet model
  for the system.  ASAS~J102322$-$3737.0 exhibited plausible evidence
  for a period increase not previously suggested; while NY~Vir, QS~Vir
  and NSVS~14256825 afforded less significant support for period change,
  providing some confirmation to earlier claims.  In other cases,
  period change was not convincingly observed; for AA~Dor and
  NSVS~07826147, previous findings of constant period were confirmed.
  This study allows us to present hundreds of new primary eclipse
  timings for these systems, and further demonstrates the value of
  wide-field high-cadence surveys like SuperWASP for the investigation
  of variable stars.}

\titlerunning{Period and amplitude variations in SuperWASP PCEBs}
\authorrunning{M.~E.~Lohr et al.}

\maketitle
\section{Introduction}

Since the discovery of planetary-mass objects around millisecond
pulsar PSR~B1257+12 \citep{wolszczan}, exoplanets have been detected
in a range of surprising environments.  Numerous hot Jupiters present
a challenge to planetary system formation models
(e.g. \citet{mayquel}); planets have been found orbiting single
members of binary and higher-order multiple star systems
(e.g. \citet{butler, anglada}); and recently a number of circumbinary
planets have been observed (e.g. \citet{doyle}).  In the last few
years, claims of planets in post-common-envelope binaries (PCEBs) have
proved especially controversial, and in this paper we aim to add to
the evidence needed to evaluate models for such systems, using
archival survey data.

\begin{figure*}
\centering
\includegraphics[width=18cm]{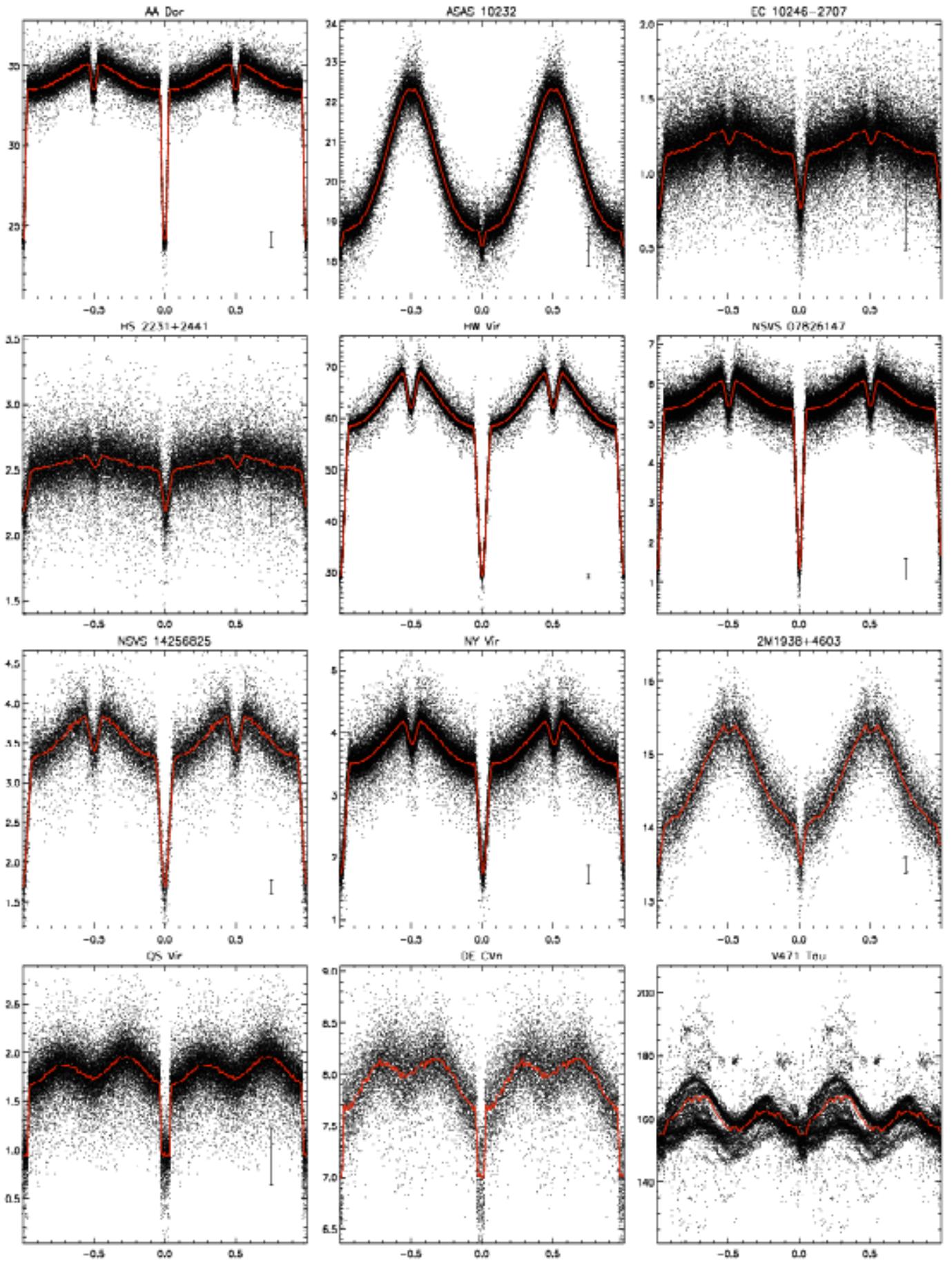}
\caption{SuperWASP light curves for 12 eclipsing PCEBs phase-folded at
  periods given in Table~\ref{pceblist}, with binned mean curves
  overplotted in red (online only).  The $x$ axes indicate phase; the
  $y$ axes SuperWASP flux in arbitrary units (pseudo-$V$ magnitudes
  are given by $15-2.5\log(flux)$.  A typical point's
  uncertainty is shown in the bottom right of each panel.}
\label{allpceblcs}
\end{figure*}

A notable class of binary star systems have passed through a phase of
common envelope evolution \citep{paczynski76}.  The details of the
process and its various possible outcomes are not yet fully understood
(for a recent review see \citet{ivanova}); however, one observed
outcome is the formation of a PCEB consisting of a hot subdwarf B
(sdB) or OB stellar core \citep{heber} or white dwarf (WD) primary,
and a low-mass main sequence star or brown dwarf companion, in a close
but detached orbital configuration.  Eclipsing PCEBs of these types
are especially valuable for understanding common envelope evolution
and subsequent system behaviours, since their parameters can be
determined with high precision.  Their photometric light curves are
also highly distinctive, often exhibiting well-defined very deep
primary eclipses and strong reflection effects (e.g. HW~Vir in
Fig.~\ref{allpceblcs}).  Together with their short orbital periods
(usually a few hours), these features facilitate accurate measurement
of timings of minimum light, and thereby the construction of observed
minus calculated (O$-$C) diagrams to reveal any changes in orbital
period over time.

\citet{zoroschreib} compiled a list of currently known eclipsing
PCEBs, including 13 with an sdB primary and 43 with a WD primary, and
noted that for five sdB systems and four WD systems apparent period
changes had been observed: a surprisingly high proportion of those
which have been well-studied over long time bases.  Many researchers
have seen these period changes as evidence for the presence of
additional massive bodies in the system: circumbinary giant
planets\footnote{Planets might also in principle be detected through
sinusoidal variations of an sdB star's pulsation period, as suggested
for isolated pulsator V391~Peg \citep{silvotti}} or brown dwarfs
e.g. \citet{lee09, beuermann10}.  The reality of such PCEB planets is
somewhat controversial.  Where multiple circumbinary planets have been
proposed in a single system, the long-term dynamical stability of
their orbits has often been questioned e.g. \citet{horner,
wittenmyer}, though the dynamical stability analyses used have also
been criticized on methodological grounds \citep{marsh}.
\citet{zoroschreib} carried out binary population syntheses which
suggested that giant planets should be rare in the progenitors of
PCEBs, leaving secondary planet formation \citep{volschow} or a
non-planetary cause such as the Applegate mechanism \citep{applegate}
as the most likely explanations for the observed period changes.

Distinguishing between different proposed architectures for
circumbinary planetary systems, and indeed determining whether planets
are plausible in PCEBs at all, relies largely upon the quality of the
eclipse timing observations: ideally, we would have a large number of
precise measurements of light curve minima, evenly covering a long
time-base.  In practice, many systems for which period changes
indicative of circumbinary planets have been claimed, fall far short
of this ideal.  Therefore, here we search the archive of the SuperWASP
(Wide Angle Search for Planets) project \citep{pollacco} for evidence
of period changes in those PCEBs from \citeauthor{zoroschreib}'s Table
1 which have been observed by SuperWASP.  The archive contains
high-cadence photometric light curves for bright sources ($V\sim$8--15
mag) over almost the whole sky, stretching back to 2004 in many cases,
and so should be capable of filling in gaps or extending the coverage
of O$-$C diagrams for many of these systems.  We have previously
observed and measured period changes in short-period main sequence
eclipsing binary candidates using SuperWASP data \citep{lohr, lohr13};
here, we develop our analytical method to improve the precision and
robustness of its period change measurements, and to search for
variations in light curve amplitude as well.  It is hoped that the
results may shed light on future investigations of this intriguing
group of eclipsing binary systems.

\section{Method}
\label{method}

\begin{table*}
\caption{PCEBs observable in SuperWASP archive}
\label{pceblist}
\centering
\begin{tabular}{l l l l l l l l}
\hline\hline\noalign{\smallskip}
System & SuperWASP ID & Type & Time base & $P$ & $\dot{P}$ & $\dot{P}$ limit & Ref. min.\\
short name & (Jhhmmss.ss$\pm$ddmmss.s) & & MM/(20)YY & (s) & (s~yr\textsuperscript{-1}) & (s~yr\textsuperscript{-1}) & (BJD-2450000)\\
\hline
\object{AA~Dor} & J053140.34$-$695302.1 & sdOB+dM/BD & 09/08--03/11 & 22597.030(3) & & 0.01 & 4738.35933(3) \\
ASAS~10232\tablefootmark{a} & J102321.90$-$373659.9 & sdB+dM & 05/06--06/11 & 12032.8839(9) & 0.0073(12) & 0.003 & 3860.199336(14)\\
\object{EC~10246$-$2707} & J102656.50$-$272256.7 & sdB+dM & 05/06--06/12 & 10239.0898(3) & 0.0012(8) & 0.003 & 3860.161068(12) \\
\object{HS~2231+2441}\tablefootmark{b} & J223421.48+245657.5 & sdB+BD(?) & 05/04--09/10 & 9554.789(2) & & 0.03 & 3150.652325(11) \\
\object{HW~Vir} & J124420.23$-$084016.8 & sdB+dM & 07/06--03/11 & 10084.5643(6)
& 0.00287(9) & 0.0003 & 3924.150763(12) \\
2M~1938+4603\tablefootmark{c} & J193832.60+460359.1 & sdB+dM &
05/04--07/10 & 10866.1147(17) & & 0.01 & 3128.537958(13) \\
\object{NSVS~07826147}\tablefootmark{d} & J153349.46+375928.2 & sdB+dM &
05/04--06/11 & 13976.9668(4) & & 0.0003 & 3128.426606(16) \\
\object{NSVS~14256825}\tablefootmark{e} & J202000.46+043756.4 & sdOB+dM & 06/06--08/11 &
9536.3263(5) & 0.0019(8) & 0.003 & 3904.675974(11) \\
\object{NY~Vir} & J133848.16$-$020149.3 & sdB+dM & 07/07--03/11 & 8727.7761(7)
& $-$0.0016(6) & 0.003 & 4307.135167(10) \\
\object{DE~CVn} & J132653.28+453246.9 & WD+dM & 05/04--03/11 & 31461.639(8) & & 0.03 & 3128.30161(4) \\
\object{QS~Vir} & J134952.07$-$131337.3\tablefootmark{f} & WD+dM & 07/07--03/11
& 13025.4555(8) & 0.007(3) & 0.01 & 4307.165601(15) \\
\object{V471~Tau} & J035024.96+171447.4 & WD+dK2 & 09/06--11/11 & 45030.05(4) &
& & \\
\hline
\end{tabular}
\tablefoot{
\tablefoottext{a}{\object{ASAS~J102322$-$3737.0}=TYC~7709$-$376$-$1.}
\tablefoottext{b}{=2MASS~J22342148+2456573.}
\tablefoottext{c}{\object{2MASS~J19383260+4603591}=TYC~3556$-$3568$-$1.}
\tablefoottext{d}{Listed as NSVS~07826247 in \citet{zoroschreib}; =2MASS~J15334944+3759282.}
\tablefoottext{e}{=2MASS~J20200045+0437564.}
\tablefoottext{f}{Archive also contains slightly poorer quality
observations of this object under the identifier 1SWASP~J134952.00$-$131336.9.}
}
\end{table*}

The SuperWASP archive was first searched for objects within 1~arcmin
of the coordinates of known bright eclipsing PCEB systems.  Matching
light curves were extracted, and checked visually for evidence of the
expected variability.  In marginal cases, and where sources
neighbouring each other on the sky exhibited a similar pattern of
variability, the custom IDL code described below was used to determine
objectively whether the eclipsing signal was detectable in the data,
or to select the source with the strongest signal amongst near
neighbours.  Once the set of usable SuperWASP eclipsing PCEBs had been
established, their Sys-Rem-corrected fluxes \citep{tamuz, mazeh}, from
a 3.5~pixel-radius aperture, formed the basis of further analysis.

\subsection{Orbital period and mean light curve determination}

Extreme outliers can often complicate the analysis of SuperWASP light
curves; here, a first pass stripped out physically-impossible data
points, and then an envelope enclosing a plausibly-relevant range of
fluxes was determined from the flux frequency distribution.

Reference orbital periods were found using a form of phase
dispersion minimization \citep{lafkin,stellingwerf}, by folding each
light curve on a range of trial periods (initially separated by 1~s),
binning the folded curves by pseudo-phase, and summing the standard
deviations of fluxes in each bin to give a total dispersion measure
per trial period.  The lowest dispersion should correspond to the best
folding, where the data points have minimal scatter about the mean
light curve shape.  The period could then be refined further by
repeating the search with smaller time steps between trial periods.
Slightly different final periods are found if different numbers of
phase bins are used to calculate dispersions; therefore, by repeating
the whole period-determination procedure with a range of binnings, a
mean reference period and an indication of its uncertainty were
determined for each object, and these values were used for the
remainder of our analysis ($P$ in Table~\ref{pceblist}).

A third stage of outlier-removal was then applied, iteratively
cleaning out points lying 4.5 standard deviations from the binned
mean flux values.  This allowed a smoother light curve template shape
to be determined for each object; the number of points used for each
template also affected its out-of-eclipse smoothness and the sharpness
of its eclipses, and this was optimized by visual inspection.

As yet the folded curves had arbitrary pseudo-phases associated with
their minima, so each deeper (primary) minimum was aligned with phase
zero in a two-step process.  First, an approximate zero-phase was
found from the bin with the lowest mean flux (this would give
inaccurate results if each bin covered a significant fraction of the
orbital period, or if the primary eclipses were flat-bottomed).  Then,
folded data points within 0.1 phases of the approximate zero point
were used to define the true zero, by mirror-folding about a number of
trial zeroes, and applying phase dispersion minimization again.  This
method has the advantage of using all the data near eclipse, rather
than just the binned means, and so is able to benefit from the long
time-base and extensive cycle-coverage of a modern time-domain
automated survey like SuperWASP.  However, it does rely on the
assumption of basically symmetric primary eclipses, like the method of
\citet{kwee}, which is still used widely with high-quality photometric
light curves covering a small number of nights.  Where eclipses are
clearly asymmetric, a minimum-flux approach to finding the zero phase
would probably be more meaningful; in these PCEBs, however, primary
eclipses were indeed highly symmetric, and so a method allowing direct
comparison between SuperWASP eclipse timings and those measured by
others using \citeauthor{kwee}'s approach was preferred here.

\subsection{Period and amplitude variation measurement}

Expected primary eclipse times (on the assumption of constant periods)
were then calculated for every cycle within the time-base covered by
SuperWASP data.  The first of these reference minima, converted to
barycentric Julian date (BJD-TDB), is given for each object in the
final column of Table~\ref{pceblist}; in combination with the given
periods, this provides SuperWASP linear ephemerides.  Using these,
each night of observed data was compared with a fitting template
covering appropriate phases, derived from the binned mean light curve
template generated earlier, interpolated by a spline curve as
necessary to match the exact times of observation.  The template could
be adjusted using three parameters: $x$-axis position (time), $y$-axis
position (flux), and scaling in the $y$-direction (amplitude of
curve).  At each fitting step, the observed curve was compared with
125 synthetic curves generated from the template by varying the three
parameters simultaneously according to a cubic grid of possible
values, and the minimum $\chi^2$ value was chosen as indicating the
best fit.  The first step had the expected $x$-$y$ location and scale
of the template as the centre of the parameter ``cube''; subsequent
steps recentred the cube on the parameter combination with the lowest
$\chi^2$ value at the previous step.  If a fitting attempt repeatedly
moved the centre to the edge of the previous cube, it was deemed not
to be converging, and was abandoned.  If the cube's centre did not
move between steps, the separation between grid points was reduced,
and the fitting step was repeated.  This continued until the
difference between adjacent steps' minimum $\chi^2$ values fell below
a critical threshold (0.001).

\begin{figure}
\resizebox{\hsize}{!}{\includegraphics{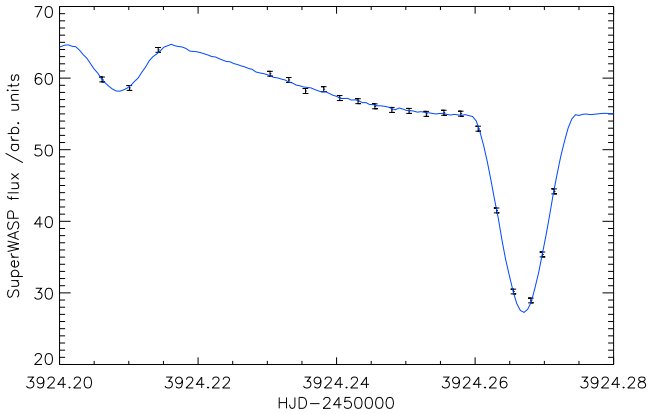}}
\caption{First night of SuperWASP observations of HW~Vir, with best
  fit overplotted (final uncertainty in timing $<2$~s).}
\label{fitex}
\end{figure}

In this way, an optimum fit between the light curve template found for
the whole data set folded at its mean (reference) period, and each
night of observed data, could be determined.  This best fit provided
an $x$-axis offset from the expected value, which corresponded to an
O$-$C value for the night as a whole, but which could also be combined
with the nearest time of calculated minimum to produce a BJD (TBD) for
that eclipse, allowing direct comparison with other published times of
minima for the same source.  Our approach here, fitting an adjustable
template light curve to the whole of a night's data (which could cover
several orbital cycles in the case of some short-period PCEBs), aimed
to take full advantage of the SuperWASP project's strengths:
long-term, numerous, fairly high-cadence observations, without being
hampered by its relatively low signal-to-noise photometry in
comparison with larger telescopes.  On many occasions, a useful time
of primary eclipse could be obtained even when the eclipse itself was
not captured by the night's observations, since the reflection effect
and shape of secondary minimum provided sufficient information for an
excellent fit (see Fig.~\ref{fitex} for an example single-night fit).

Changes in light curve amplitude could also be measured, using the
$y$-scaling parameter adjustment for the best fit.  The remaining
fixed-scale $y$-shifting parameter would track any changes in flux
level for the whole night's curve, relative to the full light curve's
out-of-eclipse mean flux; such changes might be expected to result
from varying air mass or Moon proximity on different nights, or from
instrumental noise.  Approximate starting uncertainties for each of
the three parameter values were obtained from the curvature of the
$\chi^2$ volume in the final cubic grid.

\subsection{Outlier removal and method testing}

\begin{figure*}
\centering
\includegraphics[width=18cm]{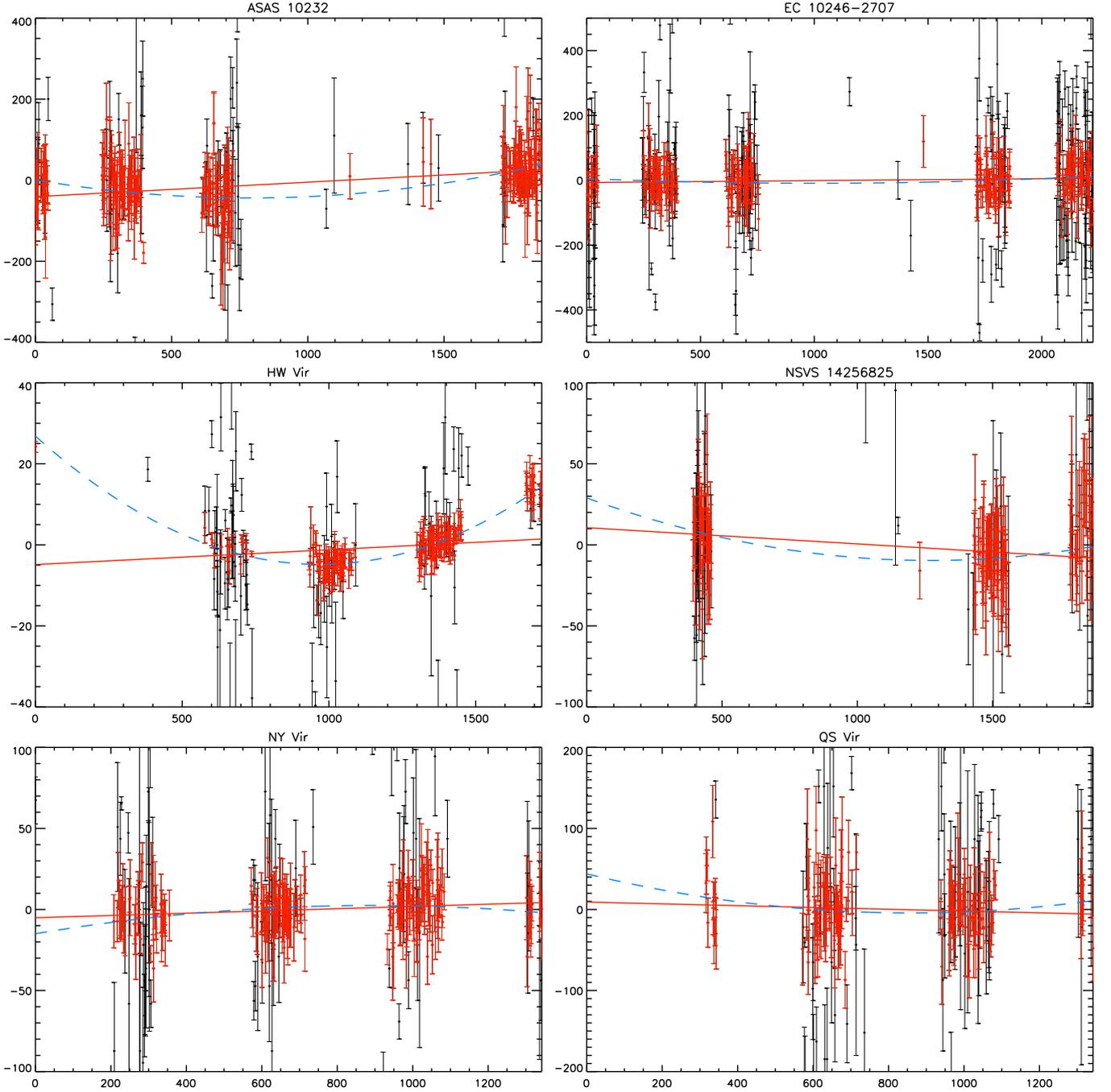}
\caption{O$-$C diagrams for six PCEBs potentially indicating period
  change, constructed using SuperWASP data only.  The $x$ axes
  indicate night count since the start of observation (since whole
  nights are fitted with template curves); the $y$ axes give O$-$C
  measurements in seconds.  Automatically-excluded outliers are in
  black and selected good minimum timings in red (colour online only);
  a small number of more extreme outliers lie outside the bounds of
  some plots.  Best linear (red solid line) and quadratic (blue dashed
  line) fits to the selected data points are overplotted.}
\label{sixocs}
\end{figure*}

After all nights had been processed, convergent results could be
plotted on three diagrams corresponding to the different fitting
parameters.  Outliers in the O$-$C diagram in particular (e.g. HW~Vir
in Fig.~\ref{sixocs}) tended to complicate the determination of period
change.  Night-by-night visual checks of the fitted data did not
suggest any underlying physical cause for short-term variations such
as spots or additional eclipses.  We may note that similar visual
checks of apparent contact eclipsing binary
1SWASP~J093010.78+533859.5, occasioned by its erratic O$-$C diagram,
revealed a second eclipsing binary \citep{lohr13}; the outlying values
here present a far more chaotic appearance, and are most probably
produced by a range of atmospheric and instrumental complicating
factors, like the outliers in SuperWASP light curves in general.

To some extent, these O$-$C outliers could be excluded by the size of
their uncertainties: some nights of data contained only a handful of
apparently erratic observations, and the resulting poor fits had large
uncertainties for their parameter values.  However, some nights
resulted in well-constrained good fits despite being obvious outliers
relative to the local O$-$C trend, so this criterion was not
sufficient.  Excluding nights with small numbers of observations would
also have removed many perfectly good values from the O$-$C diagram
(where those observations were spaced closely around the primary
minimum, for example).  Removing points on the basis that they lay
several standard deviations from the mean O$-$C value would also have
been unhelpful, since it would have removed valid points if the
underlying shape of the data set was parabolic.  It was of course not
known in advance whether a linear or quadratic fit would be
appropriate for each O$-$C diagram, and the presence of outliers could
easily change which function gave a better fit to the data set.

Therefore, an automated iterative procedure was carried out (without
any prior preference for either function) attempting linear and
quadratic fits alternately to each O$-$C diagram, and removing points
lying $>$3 standard deviations from the better fit or with
uncertainties $>$3 standard deviations larger than the mean
uncertainty size.  Sinusoidal fits were not attempted, since this
would introduce too many degrees of freedom, and since the time-base
covered appeared short enough that sinusoidal variation would show up
as approximately quadratic in any case.  The plot of amplitude
variation was also used to exclude extreme outliers in that dimension;
however the absolute flux variation plot was not used, since sudden
and substantial variations in that dimension appeared entirely
physically plausible.  If the $\chi^2$ value of the better fit ever
fell below 1, the uncertainties of the remaining points were rescaled
accordingly.  The process halted when no further points needed to be
removed\footnote{All eclipse timings, including those removed by this
process, are available in the electronic version of this
paper.}.

Following this procedure, period change was either supported, if a
quadratic function gave a better fit to the remaining data points in
the O$-$C diagram, or unsupported, if a linear function gave a better
fit.  (A linear fit with slope significantly different from zero would
also arise if the period used were too long or too short, though
irregular or sparse coverage of the time base could also produce a
non-zero gradient even with an accurate period.)  However, some cases
of apparent period change were only marginally supported, in that the
best linear fit produced a (modified) $\chi^2$ value only slightly
higher than the best quadratic fit.  Since the points' uncertainties
had been adjusted during the process of outlier removal, it was not
clear how large a difference in $\chi^2$ values would be required to
indicate e.g. a 95\% confidence level in a measurement of period
change.  In particular, there seemed no reason to believe that the
same level would be valid in all cases.

Therefore, tests using synthetic light curves were carried out to
determine the reliability of the program.  Each object's mean
(template) light curve was used as the basis for generating a large
number of ``background'' synthetic curves, and the time sampling and
point uncertainties of the original light curve were applied to each
synthetic curve.  Each flux value was then perturbed randomly
according to a normal distribution with standard deviation equal to
the corresponding original data point's uncertainty.  Correlations
between observations made on the same night in SuperWASP light curves
were accounted for by determining the mean residual flux of each
night's observations relative to the template, and adding one of these
values, chosen at random, to each night's fluxes in the synthetic
curve.  Histograms of the final synthetic curves' residual fluxes,
relative to their mean curves, followed approximately Gaussian
distributions (like the original objects), but with slightly greater
widths i.e. the synthetic curves had slightly larger uncertainties
than the real light curves.  No period change was included in the
synthetic curves.  They were then processed by our code in exactly the
same way as the real light curves, to see what proportion of them
produced false positives, and how large the difference between best
linear and quadratic fit $\chi^2$ values was.  This allowed us to
distinguish between statistically significant and non-significant
period changes.

A similar approach was used to check the sensitivity of the program to
genuine period change.  Synthetic curves were generated as before,
with the characteristics of the test objects; here, however, steady
period change was included, with a known sign and magnitude.  Our code
was then run on the synthetic curves, to determine lower limits of
detectability for each system i.e.  how rapid a period change would
need to be in order to be reliably detected and accurately measured
using SuperWASP archive data.

\section{Results}

Of \citeauthor{zoroschreib}'s collected eclipsing PCEBs, twelve were
bright enough to have usable observations in the SuperWASP archive, of
which nine were HW~Vir-type systems (sdB or sdOB primary with an M
dwarf or brown dwarf companion), and three contained a WD with a
low-mass main sequence companion (Table~\ref{pceblist}).  In the case
of QS~Vir, two nearby sources fell within the SuperWASP aperture,
resulting in a pair of very similar archive light curves containing
the eclipsing signal of the same system.  One curve was slightly
brighter and had a larger amplitude, and was selected for further
analysis here.

Orbital periods were determined for the twelve objects by the method
described in Sect.~\ref{method}, accurate to between 7 and 9
significant figures (see Table~\ref{pceblist}, which also gives the
date ranges during which they were observed by SuperWASP).  Their
light curves, phase-folded using these periods, are shown in
Fig.~\ref{allpceblcs}.  All exhibit a strong reflection effect; in
ASAS~10232 and 2M~1938+4603 this dominates the light curve shape.  The
other systems all show deep, well-defined primary eclipses, which are
flat-bottomed in the cases of the WD systems and AA~Dor.  V471~Tau
exhibited extreme short-term variability in light curve shape and
amplitude, which prevented further analysis of possible period changes
since no typical template curve could be determined for it.  It
therefore plays no further part in this study, though an individual
customized analysis of its SuperWASP archive data might yield useful
results in future.

The remaining eleven objects were searched for evidence of period
change, and a selection of their O$-$C diagrams are given in
Fig.~\ref{sixocs}.  In six cases, the best fit to the data following
outlier removal was quadratic (ASAS~10232, EC~10246$-$2707, HW~Vir,
NSVS~14256825, NY~Vir and QS~Vir), so the significance of the period
change implied was tested for these objects.  HW~Vir exhibited highly
significant period increase over the observed time base ($p$-value of
0.002 i.e. 0.2\% of ``background'' tests provided equally strong or
stronger evidence for period change, purely by chance).  ASAS~10232
and NY~Vir showed very plausible evidence for period change
($p$-values of 0.02 and 0.06 respectively).  Even considering that 11
trials were run, we would expect to find three or more results with
$p\le$0.06 by accident only $\sim$2.5\% of the time.  QS~Vir,
NSVS~14256825 and EC~10246$-$2707 provided increasingly weak and
non-significant support for period change ($p$-values of 0.22, 0.28
and 0.36 respectively); however, we note that the direction and
approximate magnitude of the changes suggested for QS~Vir and
NSVS~14256825 accord with the findings of other researchers (see
Subsects.~\ref{discussqsvir} and \ref{discussnsvs} below).  The other
five objects did not show evidence of period change over the time
bases considered.

The full set of SuperWASP light curves had widely-varying
sensitivities to genuine change: for DE~CVn changes of up to
0.03~s~yr\textsuperscript{-1} would not have been detectable, while
for NSVS~07826147, any changes would have to be slower than
0.0003~s~yr\textsuperscript{-1} to be missed.  Table~\ref{pceblist}
gives the measured period changes ($\dot{P}$) and the limits of
expected period change detectability ($\dot{P}$ limit).
Tables~\ref{aadordat}--\ref{qsvirextra} (online only) give the times
of minima for the eleven objects, in BJD (BDT).

No clear evidence of change in light curve amplitudes was observed,
though there was a possible suggestion of curvature in ASAS~10232's
amplitude-time diagram which might repay further investigation.

\section{Discussion}
\label{discussion}

\begin{figure*}
\centering
\includegraphics[width=18cm]{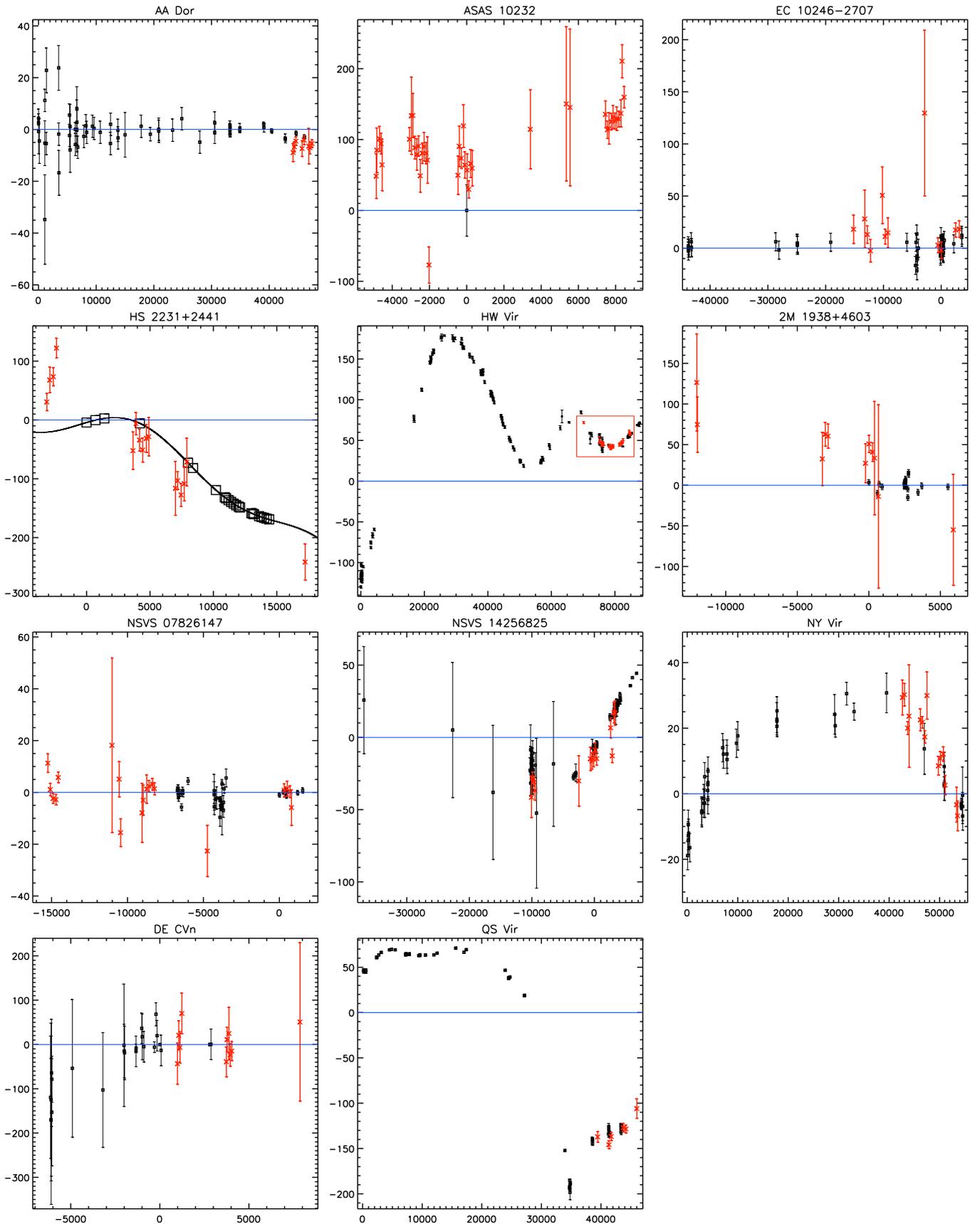}
\caption{O$-$C diagrams for eleven PCEBs according to ephemerides
  given in text, including previously published eclipse timings (black
  squares) and new binned SuperWASP timings (red crosses).  The $x$
  axes indicate cycle count; the $y$ axes O$-$C values in seconds.
  For HW~Vir, the region containing new SuperWASP values is surrounded
  by a red rectangle for clarity.  For HS~2231+2441, larger black
  squares indicate the approximate locations of the unpublished
  observations of \citet{qian10}, with their fitted curve overplotted
  as a solid line.}
\label{allfullocs}
\end{figure*}

For easier comparison with others' findings, previously-published
eclipse timings were collected for each object, and converted to
BJD (TDB) where
necessary\footnote{http://astroutils.astronomy.ohio-state.edu/time/hjd2bjd.html
(see also \citet{eastman}).}.  O$-$C diagrams were then compiled
relative to a recent or widely-used ephemeris, and the new SuperWASP
values were included after seasonal binning to improve the
clarity of the trends they indicate (Fig.~\ref{allfullocs}).
Objects are discussed individually.

\subsection{AA~Dor}

AA~Dor was discovered, identified as an eclipsing sdB binary, and
given an initial solution in a series of papers by
\citet{kilkenny78,kilkenny79,kilkenny81}, and eclipse timings have
been published for it covering the period 1977--2010
\citep{kilkenny86,kilkenny91,kilkenny00,kilkenny11}.  The ephemeris
used here is from \citet{kilkenny11}:
\begin{equation}\mathrm{BJD} \quad 2443196.34925+0.2615397362E.\end{equation}
They conclude that a linear ephemeris is sufficient to explain the
observations to date, and the addition of our partially-overlapping
timings, extending coverage to March 2011, does not contradict this.
They suggest that any change greater than about 3$\times
10^{-7}$~s~yr\textsuperscript{-1} would be ruled out.  AA~Dor is
notable within the set of objects because, as \citeauthor{zoroschreib}
point out, `it is so far the only PCEB with continuous high-precision
eclipse time measurements that does not show any signs of apparent
period variations'.

\subsection{ASAS~10232}

The discovery paper for ASAS~10232 as an eclipsing sdB binary is
\citet{schaffenroth}, which also provides an orbital solution, but
only a single time of minimum obtained from Carnes Hill Observatory
BVI light curves from March 2008: HJD~2454538.99689$\pm$0.00042 (or
possibly 0.000042).  They determine a period from the first three
years of SuperWASP archived observations, covering May 2006--January
2009; since our data points used here form a superset of their data,
extending coverage to June 2011, we prefer our period
(Table~\ref{pceblist}) with their cycle zero to form the ephemeris:
\begin{equation}\mathrm{BJD} \quad 2454538.99765+0.139269490E.\end{equation}
\citeauthor{schaffenroth} also measured individual eclipse times from
their SuperWASP data, by a similar method to that used here (fitting a
mean light curve to night-by-night data), to produce an O$-$C diagram
(their Fig.~7), to which they fit a downward-opening parabola.  On the
basis of this they suggest a possible period decrease in the system,
but note that it requires more observations to confirm.  The fitting
method used here is able to benefit from a longer time-base and hence
a better-defined mean light curve; thus we have apparently been able
to measure the times of eclipse more precisely from the same source of
data.  Although their point uncertainties are not shown, the majority
of their O$-$C values seem to fall within $\sim$400~s of the fitted
curve; ours (Fig.~\ref{sixocs}) fall mostly within $\sim$100~s of
the preferred quadratic fit.  After binning every two weeks' O$-$C
values together (Fig.~\ref{allfullocs}) the trend is even more
clearly indicated: over six years, the system appears to be
increasing, rather than decreasing in period.  Of course, the
variation may be more complex e.g. sinusoidal, on a still longer
timescale, and further independent observations are needed to help
clarify the behaviour of this system.

\subsection{EC~10246$-$2707}

Although previously known to contain an sdB star, EC~10246$-$2707 was
not described as an eclipsing binary until \citet{barlow}, which also
estimates the system parameters and provides eclipse timings between
February 1997 and June 2012.  We use their ephemeris:
\begin{equation}\mathrm{BJD} \quad 2455680.562160+0.1185079936E.\end{equation}
They find no evidence for period change on the basis of their data,
and determine a limit on detectability of change of
0.0003~s~yr\textsuperscript{-1}.  Our results here support this
non-detection of change, and help to fill a gap in their coverage of
the system's behaviour during 2006--2008.

\subsection{HS~2231+2441}

The discovery paper for HS~2231+2441 is \citet{ostensen07}, who
determine system parameters and provide the ephemeris:
\begin{equation}\mathrm{BJD} \quad 2453522.669493+0.1105880E.\end{equation}
\citet{qian10} describe observations of the system between 2005 and
2009, and provide an O$-$C diagram (their Fig.~4), which they fit with
a function including both a quadratic and a sinusoidal term.
Therefore, they suggest the presence of a secular decrease in orbital
period, associated with magnetic braking, and a tertiary companion
responsible for the sinusoidal variation.  Since their times of minima
do not appear to have been published yet, we compare our SuperWASP
eclipse timings with their fitted curve, and estimate the cycle
numbers of their observations from their O$-$C diagram; their data
values are placed on the fitted curve despite exhibiting some scatter
about it (Fig.~\ref{allfullocs}).  Although fairly close to their fit
during 2006 and 2007, our observations do not strongly support it
outside their original data range i.e. during 2004 and 2012, and a
linear function might provide a better fit to the full data set.  A
straight-line fit with negative slope would be expected in an O$-$C
diagram if the period used to construct it were too long; we note that
\citeauthor{ostensen07}'s period, based on just three nights of
observations during June and September 2005, is fractionally longer
than ours (0.11058784~d), and quoted to lower precision, and this may
be a cause for the apparent long-term downward trend seen here.

\subsection{HW~Vir}

The prototype for eclipsing sdB binaries, HW~Vir was discovered by
\citet{menzies}, and its times of minima were documented between 1984
and 2002 by a group at the South African Astronomical Observatory
\citep{marang, kilkenny91, kilkenny94, kilkenny00, kilkenny03}, who
also studied AA~Dor and NY~Vir over many years.  Following
\citet{beuermann12b}, we include their eclipse timings in
Fig.~\ref{allfullocs} along with others having a quoted error
$\leq$0.0001~d \citep{wood, lee09, brat}, and
\citeauthor{beuermann12b}'s own results, up to February 2012.  For
clearer comparison of our results with the recent models of
\citeauthor{lee09} and \citeauthor{beuermann12b}, we use their
ephemeris:
\begin{equation}\mathrm{BJD} \quad 2445730.55803+0.1167195E.\end{equation}
\citeauthor{lee09} interpreted the O$-$C diagram up to 2009 (epoch
$\sim$76000; their Fig.~5 top panel) as the sum of a downward-opening
parabola (secular period decrease caused by magnetic braking) and two
sinusoidal terms (LITE associated with two substellar circumbinary
companions).  However, \citeauthor{beuermann12b} pointed out that the
proposed companions' orbits crossed, indicating a probable
near-encounter or collision within 2000~y; moreover, the O$-$C values
after 2009 diverge substantially from \citeauthor{lee09}'s fit,
curving upwards rather than following the proposed quadratic decline.
They argue for an alternative model without the long-term period
decrease, and involving two circumbinary low-mass objects in orbits
which they found to be stable for more than $10^8$~y.  We note that
our new SuperWASP eclipse timings, covering July 2006 to March 2011,
strongly support \citeauthor{beuermann12b}'s model over that of
\citeauthor{lee09}, in that a significant period increase is clear,
and several previously undocumented parts of the general trend during
this time are now well covered.  HW~Vir is also the system in which
the contribution of SuperWASP archival data is most readily apparent:
some 180 primary eclipse timings could be measured with uncertainties
below 0.00006~d, covering about six years.

\subsection{2M~1938+4603}

2M~1938+4603, a \emph{Kepler}-field object \citep{borucki} known to
contain an sdB star, was observed to possess shallow primary and
secondary eclipses, in addition to its substantial reflection effect,
by \citet{ostensen10} (who also discovered HS~2231+2441).  On the
basis of 13 supplementary ground-based timings covering June 2008--May
2010, they provide the ephemeris
\begin{equation}\mathrm{BJD} \quad 2454640.86416+0.12576530E.\end{equation}
Our SuperWASP timings extend coverage back to May 2004, and although
they are individually not very precise, when binned together they
suggest a long-term negative linear trend.  As with HS~2231+2441, we
suspect \citeauthor{ostensen10}'s period is fractionally too long
(ours is 0.12576522~d), creating a downward slope in the O$-$C diagram
(their timing uncertainties may also be underestimated, given the
scatter of their observations about the mean).  Allowing for this, our
data set does not seem to suggest any period change in this system.
It is interesting to note that \citeauthor{ostensen10} also provide 77
extremely precise consecutive eclipse timings from \emph{Kepler}
observations (around epoch 2500); when the full continuous space-based
light curve for this object is made available, it should be possible
to determine whether 2M~1938+4603 is undergoing period variations with
unprecedented confidence and precision.

\subsection{NSVS~07826147}

NSVS~07826147 was discovered by \citet{kelley}, and primary eclipse
timings have been published for it by \citet{for, liying, backhaus}.
We use \citeauthor{backhaus}'s ephemeris to construct our O$-$C
diagram:
\begin{equation}\mathrm{BJD} \quad 2455611.926580+0.1617704531E.\end{equation}
No period change has been claimed yet for this system, though the
previously-published timings only covered February 2008--October 2011;
the addition of our SuperWASP timings extends the coverage back to May
2004, and provides stronger support for a constant orbital period.
Indeed, our results suggest an upper limit on any period variation of
about 0.0003~s~yr\textsuperscript{-1}.

\subsection{NSVS~14256825}
\label{discussnsvs}

NSVS~14256825 was identified as an eclipsing sdB binary by
\citet{wils}, who published some eclipse timings; others have been
provided by \citet{kilkenny12, beuermann12a, almeida13}.
\citet{qian10} report observations of the system since 2006, and claim
evidence for a cyclic variation, but have not yet published supporting
timing measurements.  Here, we use the ephemeris of the most recent
analysis of the system, \citet{hinse}:
\begin{equation}\mathrm{BJD} \quad 2455408.74454+0.11037411E.\end{equation}
On the basis of very similar O$-$C variations, \citet{beuermann12a}
argue for a single circumbinary low-mass companion in an elliptical
orbit, while \citet{almeida13} prefer a two-planet model.
\citet{hinse}, however, find that the data are insufficient to
constrain any particular one-planet model, and provide no convincing
evidence for a second circumbinary companion.  Unfortunately, while
our new timings are quite consistent with previous measurements, and
independently support period increase over June 2006--August 2011,
they do not add much to the coverage or clarify the longer-term trends
of period variation for this particular system.

\subsection{NY~Vir}

\citet{kilkenny98} published the discovery paper for NY~Vir, and
provided eclipse timings and an ephemeris for it between 1996 and 2010
\citep{kilkenny00, kilkenny11}.  Additional times are given in
\citet{camurdan} and \citet{qian12}; the latter also provides the
revised ephemeris:
\begin{equation}\mathrm{BJD} \quad 2450223.362213+0.1010159673E.\end{equation}
A steady period decrease was observed in the O$-$C diagram by
\citet{kilkenny11, camurdan, qian12}, and is independently supported
by our new SuperWASP timings.  \citeauthor{qian12} argue that this is
unlikely to be caused by the Applegate mechanism, gravitational
radiation or magnetic braking, due to its magnitude and the fully
convective nature of the stars, and suggest instead that it is part of
a long-term ($>$15~y) cyclic variation associated with a circumbinary
planet; furthermore, they claim that the O$-$C diagram provides
evidence for a shorter-period fourth body in the system.

\subsection{DE~CVn}

DE~CVn was identified as an X-ray source in the ROSAT catalogue
\citep{voges}, and as an eclipsing binary containing a WD by
\citet{robb}, who also published several times of minima.  Other
timings are provided by \citet{besselaar, tas, parsons}, and
\citeauthor{parsons} also give the ephemeris we use in
Fig.~\ref{allfullocs}:
\begin{equation}\mathrm{BJD} \quad 2452784.554043+0.3641393156E.\end{equation}
Although the previously-published eclipse timings of DE~CVn cover
1997--2006, \citeauthor{parsons} feel that most are too uncertain to
allow any claims regarding period change to be made.  Our new timings
extend coverage to March 2011, and although they also have large
uncertainties, we may note at least that the whole O$-$C diagram is
fully consistent with a constant period for this system, over about 14
years.

\subsection{QS~Vir}
\label{discussqsvir}

QS~Vir was discovered and later identified as an eclipsing WD binary
by \citet{kilkenny97, odonoghue}.  They and \citet{kawka, qian10b,
  parsons} provide eclipse timings for it, and here we use the
ephemeris of \citeauthor{parsons}:
\begin{equation}\mathrm{BJD} \quad 2448689.64062+0.150757525E.\end{equation}
(\citet{almeida11} also refer to new timings for the system, but have
not yet published them.)  The substantial period changes evident in
Fig.~\ref{allfullocs} are demonstrated by \citeauthor{parsons} to be
an order of magnitude too large to be caused by the Applegate
mechanism; however, they are also doubtful about the plausibility of a
third body in the system, while noting that it ``remains the only
mechanism able to produce such a large period variation''.  The data
set available to them covered April 1992--February 2010;
\citet{almeida11} add a few more points extending it to August 2010,
and argue on this basis for a system containing two circumstellar
low-mass bodies.  Our partly-overlapping new timings extend the time
base to March 2011, and provide independent, if weak, support for a
recent increase in QS~Vir's orbital period.

\section{Conclusions}

Twelve PCEBs with observations covering between three and seven years
in the SuperWASP archive were analysed here for evidence of period
and/or light curve amplitude change, potentially indicating the
presence of circumbinary planets.  Their periods were found to high
precision, agreeing very closely with those found in previous studies.
Hundreds of primary eclipse timings were also determined for the
objects, in many cases for previously unobserved epochs, and are made
available in the electronic version of this article, for future study
of these systems' period variations.

Period changes found in much previous work were strongly confirmed
here for HW~Vir, as was the stability of the periods of AA~Dor and
NSVS~07826147.  New eclipse timings of NSVS~14256825, NY~Vir and
QS~Vir, previously suggested as hosts for third bodies, provided some
support for period change, while claims of period variations for
HS~2231+2441 were not supported by our data.  V471~Tau could not be
analysed for period variations due to its dramatic and apparently
irregular amplitude changes.  For 2M~1938+4603 and DE~CVn, previously
published eclipse timings had not been sufficient to make strong
claims; we found no plausible evidence for period changes in these
systems.  However, for ASAS~10232, our data provided fairly strong
evidence for period increase between May 2006 and June 2011, and
perhaps for systematic amplitude changes as well, which might suggest
this system as a further candidate for containing a circumbinary third
body.

\begin{acknowledgements}
The WASP project is currently funded and operated by Warwick
University and Keele University, and was originally set up by Queen's
University Belfast, the Universities of Keele, St. Andrews and
Leicester, the Open University, the Isaac Newton Group, the Instituto
de Astrofisica de Canarias, the South African Astronomical Observatory
and by STFC.  This work was supported by the Science and Technology
Funding Council and the Open University.
\end{acknowledgements}

\bibliographystyle{aa}
\bibliography{reflist}

\Online
\begin{appendix}
\section{Primary eclipse timings for 11 PCEBs}
\begin{table*}
\caption{Selected good SuperWASP times of minimum light for AA~Dor.}
\label{aadordat}
\centering
\begin{tabular}{l l | l l | l l | l l}
\hline\hline
Min. time & Uncertainty & Min. time & Uncertainty & Min. time & Uncertainty & Min. time & Uncertainty \\
(BJD$-$2450000) & (d) & (BJD$-$2450000) & (d) & (BJD$-$2450000) & (d) & (BJD$-$2450000) & (d) \\
\hline
4738.620770 & 0.000051 & 4844.544438 & 0.000185 & 5157.607588 & 0.000170 & 5458.639884 & 0.000145 \\
4743.590065 & 0.000066 & 4846.636483 & 0.000161 & 5162.576787 & 0.000109 & 5477.732064 & 0.000071 \\
4752.482619 & 0.000123 & 4847.682751 & 0.000149 & 5163.622565 & 0.000278 & 5486.624578 & 0.000085 \\
4758.497923 & 0.000123 & 4848.729319 & 0.000176 & 5171.730615 & 0.000111 & 5487.670627 & 0.000228 \\
4763.728689 & 0.000141 & 4849.513679 & 0.000128 & 5174.607450 & 0.000156 & 5488.716813 & 0.000196 \\
4765.559304 & 0.000224 & 4850.559879 & 0.000099 & 5178.530920 & 0.000105 & 5496.562827 & 0.000141 \\
\hline
\end{tabular}
\end{table*}

\begin{table*}
\caption{Additional SuperWASP times of minimum light for AA~Dor (excluded from analysis here).}
\centering
\begin{tabular}{l l | l l | l l | l l}
\hline\hline
Min. time & Uncertainty & Min. time & Uncertainty & Min. time & Uncertainty & Min. time & Uncertainty \\
(BJD$-$2450000) & (d) & (BJD$-$2450000) & (d) & (BJD$-$2450000) & (d) & (BJD$-$2450000) & (d) \\
\hline
4744.635965 & 0.000355 & 4868.606117 & 0.000263 & 5217.500722 & 0.000341 & 5521.668880 & 0.000410 \\
4746.729045 & 0.000631 & 4872.529322 & 0.000621 & 5221.684268 & 0.000600 & 5526.639878 & 0.000210 \\
4749.605491 & 0.000402 & 4877.497868 & 0.000276 & 5225.607145 & 0.000565 & 5544.685790 & 0.000837 \\
4750.651432 & 0.000403 & 4882.729261 & 0.000352 & 5231.625391 & 0.000597 & 5550.701857 & 0.000310 \\
4751.697482 & 0.000239 & 4888.482480 & 0.000420 & 5234.499931 & 0.000447 & 5551.479992 & 0.000122 \\
4753.536052 & 0.000368 & 4894.498874 & 0.000383 & 5235.545218 & 0.000546 & 5566.635454 & 0.000250 \\
\hline
\end{tabular}
\end{table*}

\begin{table*}
\caption{Selected good SuperWASP times of minimum light for ASAS~10232.}
\centering
\begin{tabular}{l l | l l | l l | l l}
\hline\hline
Min. time & Uncertainty & Min. time & Uncertainty & Min. time & Uncertainty & Min. time & Uncertainty \\
(BJD$-$2450000) & (d) & (BJD$-$2450000) & (d) & (BJD$-$2450000) & (d) & (BJD$-$2450000) & (d) \\
\hline
3860.615868 & 0.000572 & 4178.569482 & 0.000565 & 4525.489585 & 0.000442 & 5629.479054 & 0.000503 \\
3862.566685 & 0.000645 & 4179.544020 & 0.000442 & 4526.464182 & 0.000530 & 5631.568590 & 0.000444 \\
3863.541688 & 0.000691 & 4180.519400 & 0.000627 & 4527.578047 & 0.000596 & 5632.543070 & 0.000536 \\
3864.516806 & 0.000794 & 4181.493764 & 0.000510 & 4528.553398 & 0.000765 & 5633.518073 & 0.000535 \\
3865.491316 & 0.000581 & 4183.582981 & 0.000408 & 4536.492398 & 0.000420 & 5634.493568 & 0.000503 \\
3866.466173 & 0.000854 & 4184.557867 & 0.000527 & 4539.555544 & 0.000539 & 5635.467555 & 0.000576 \\
\hline
\end{tabular}
\end{table*}

\begin{table*}
\caption{Additional SuperWASP times of minimum light for ASAS~10232 (excluded from analysis here).}
\centering
\begin{tabular}{l l | l l | l l | l l}
\hline\hline
Min. time & Uncertainty & Min. time & Uncertainty & Min. time & Uncertainty & Min. time & Uncertainty \\
(BJD$-$2450000) & (d) & (BJD$-$2450000) & (d) & (BJD$-$2450000) & (d) & (BJD$-$2450000) & (d) \\
\hline
3868.564326 & 0.001873 & 4165.486419 & 0.000905 & 4556.545406 & 0.002321 & 4928.534811 & 0.000553 \\
3869.530392 & 0.001224 & 4166.454691 & 0.000737 & 4562.536199 & 0.000536 & 4957.504955 & 0.001642 \\
3872.456097 & 0.001173 & 4182.468361 & 0.000525 & 4567.543982 & 0.002119 & 5228.522572 & 0.001158 \\
3882.482470 & 0.000500 & 4226.472965 & 0.000651 & 4571.448402 & 0.000674 & 5229.505583 & 0.001122 \\
3895.573323 & 0.001048 & 4230.521558 & 0.003851 & 4577.578814 & 0.001199 & 5283.534485 & 0.001011 \\
3907.553285 & 0.000616 & 4248.484538 & 0.001235 & 4584.542607 & 0.000579 & 5311.520806 & 0.001114 \\
3922.449259 & 0.000461 & 4249.457161 & 0.001116 & 4585.515898 & 0.000375 & 5340.495127 & 0.000946 \\
4122.583060 & 0.001458 & 4253.497891 & 0.001175 & 4586.489740 & 0.001794 & 5576.555285 & 0.001054 \\
4128.578321 & 0.000717 & 4254.474170 & 0.001070 & 4592.471190 & 0.000876 & 5581.569915 & 0.000983 \\
4131.495959 & 0.000988 & 4488.582241 & 0.001565 & 4594.566442 & 0.001530 & 5582.550952 & 0.001693 \\
4133.447008 & 0.001894 & 4490.534045 & 0.000759 & 4600.558743 & 0.001247 & 5585.470853 & 0.001498 \\
4135.533729 & 0.001504 & 4491.516301 & 0.001538 & 4601.536183 & 0.001099 & 5688.531029 & 0.000556 \\
4137.484895 & 0.000839 & 4509.470693 & 0.000345 & 4604.455040 & 0.001967 & 5708.445522 & 0.000506 \\
4138.459433 & 0.001763 & 4513.511249 & 0.001027 & 4605.569776 & 0.001499 &  &  \\
4162.551315 & 0.001128 & 4531.478057 & 0.001354 & 4609.466420 & 0.000938 &  &  \\
4163.517845 & 0.002328 & 4540.530865 & 0.000811 & 4615.455821 & 0.000857 &  &  \\
\hline
\end{tabular}
\end{table*}

\begin{table*}
\caption{Selected good SuperWASP times of minimum light for EC~10246$-$2707.}
\centering
\begin{tabular}{l l | l l | l l | l l}
\hline\hline
Min. time & Uncertainty & Min. time & Uncertainty & Min. time & Uncertainty & Min. time & Uncertainty \\
(BJD$-$2450000) & (d) & (BJD$-$2450000) & (d) & (BJD$-$2450000) & (d) & (BJD$-$2450000) & (d) \\
\hline
3860.636779 & 0.000880 & 4233.579709 & 0.000326 & 5579.593446 & 0.000679 & 5946.492675 & 0.000573 \\
3862.531425 & 0.000758 & 4237.489806 & 0.000683 & 5580.542102 & 0.000995 & 5952.538087 & 0.000375 \\
3863.478798 & 0.000746 & 4239.505133 & 0.000558 & 5588.482334 & 0.000723 & 5953.486521 & 0.000510 \\
3864.544530 & 0.000810 & 4247.563479 & 0.001007 & 5589.548116 & 0.000368 & 5954.552204 & 0.000499 \\
3865.493137 & 0.000934 & 4254.555450 & 0.000606 & 5590.496378 & 0.000370 & 5955.500565 & 0.000378 \\
3870.589178 & 0.000897 & 4257.518051 & 0.000445 & 5591.562357 & 0.000584 & 5956.566840 & 0.000308 \\
\hline
\end{tabular}
\end{table*}

\begin{table*}
\caption{Additional SuperWASP times of minimum light for EC~10246$-$2707 (excluded from analysis here).}
\centering
\begin{tabular}{l l | l l | l l | l l}
\hline\hline
Min. time & Uncertainty & Min. time & Uncertainty & Min. time & Uncertainty & Min. time & Uncertainty \\
(BJD$-$2450000) & (d) & (BJD$-$2450000) & (d) & (BJD$-$2450000) & (d) & (BJD$-$2450000) & (d) \\
\hline
3866.560302 & 0.001223 & 4246.496808 & 0.001028 & 5586.579973 & 0.000292 & 5963.558515 & 0.001199 \\
3867.509353 & 0.001730 & 4248.511740 & 0.001338 & 5600.566495 & 0.000777 & 5969.482532 & 0.000925 \\
3868.573160 & 0.002266 & 4249.578707 & 0.001504 & 5604.486196 & 0.000305 & 5975.525846 & 0.000629 \\
3869.523594 & 0.002122 & 4253.489865 & 0.001283 & 5624.510638 & 0.000263 & 5976.474108 & 0.001557 \\
3872.482936 & 0.001579 & 4485.529295 & 0.001459 & 5635.531385 & 0.001154 & 5977.545617 & 0.000957 \\
3882.561546 & 0.000468 & 4488.491402 & 0.001536 & 5638.488555 & 0.000996 & 5980.495281 & 0.000565 \\
\hline
\end{tabular}
\end{table*}

\begin{table*}
\caption{Selected good SuperWASP times of minimum light for HS~2231+2441.}
\centering
\begin{tabular}{l l | l l | l l | l l}
\hline\hline
Min. time & Uncertainty & Min. time & Uncertainty & Min. time & Uncertainty & Min. time & Uncertainty \\
(BJD$-$2450000) & (d) & (BJD$-$2450000) & (d) & (BJD$-$2450000) & (d) & (BJD$-$2450000) & (d) \\
\hline
3163.478486 & 0.000914 & 3233.482981 & 0.001065 & 3975.527061 & 0.000868 & 4316.470536 & 0.001005 \\
3166.466569 & 0.001421 & 3235.473193 & 0.000855 & 3978.512103 & 0.001095 & 4318.458905 & 0.001191 \\
3169.452717 & 0.000997 & 3236.469036 & 0.001033 & 3979.507946 & 0.000982 & 4319.453274 & 0.000914 \\
3170.557397 & 0.000674 & 3237.462852 & 0.001251 & 3980.503790 & 0.001318 & 4320.450131 & 0.000277 \\
3172.547333 & 0.000440 & 3238.458512 & 0.001142 & 3990.455588 & 0.001030 & 4321.556286 & 0.001119 \\
3173.543545 & 0.000636 & 3239.454539 & 0.000816 & 3991.451247 & 0.000887 & 4323.544471 & 0.000982 \\
\hline
\end{tabular}
\end{table*}

\begin{table*}
\caption{Additional SuperWASP times of minimum light for HS~2231+2441 (excluded from analysis here).}
\centering
\begin{tabular}{l l | l l | l l | l l}
\hline\hline
Min. time & Uncertainty & Min. time & Uncertainty & Min. time & Uncertainty & Min. time & Uncertainty \\
(BJD$-$2450000) & (d) & (BJD$-$2450000) & (d) & (BJD$-$2450000) & (d) & (BJD$-$2450000) & (d) \\
\hline
3152.531212 & 0.000471 & 3215.456610 & 0.001051 & 3902.533011 & 0.000340 & 4339.475937 & 0.003997 \\
3153.529820 & 0.002944 & 3216.452453 & 0.001850 & 3922.562218 & 0.002503 & 4340.463118 & 0.002901 \\
3154.529442 & 0.000688 & 3217.557595 & 0.001603 & 3954.515096 & 0.001936 & 4360.483569 & 0.002026 \\
3156.512005 & 0.007134 & 3218.553991 & 0.002251 & 3960.488314 & 0.002722 & 4364.465837 & 0.001465 \\
3157.509507 & 0.001609 & 3219.547807 & 0.001698 & 3963.471052 & 0.001624 & 4371.544933 & 0.001986 \\
3158.508576 & 0.000957 & 3220.544572 & 0.001743 & 3976.521890 & 0.001988 & 4372.542067 & 0.002438 \\
\hline
\end{tabular}
\end{table*}

\begin{table*}
\caption{Selected good SuperWASP times of minimum light for HW~Vir.}
\centering
\begin{tabular}{l l | l l | l l | l l}
\hline\hline
Min. time & Uncertainty & Min. time & Uncertainty & Min. time & Uncertainty & Min. time & Uncertainty \\
(BJD$-$2450000) & (d) & (BJD$-$2450000) & (d) & (BJD$-$2450000) & (d) & (BJD$-$2450000) & (d) \\
\hline
3924.617921 & 0.000016 & 4923.503048 & 0.000022 & 5237.478551 & 0.000039 & 5319.532357 & 0.000022 \\
4501.562165 & 0.000044 & 4924.553561 & 0.000023 & 5239.579447 & 0.000024 & 5321.516619 & 0.000045 \\
4524.555874 & 0.000019 & 4926.537724 & 0.000042 & 5240.513236 & 0.000023 & 5322.567106 & 0.000026 \\
4527.473857 & 0.000021 & 4927.471517 & 0.000039 & 5242.497525 & 0.000022 & 5324.551320 & 0.000025 \\
4534.476984 & 0.000035 & 4935.525125 & 0.000036 & 5245.532232 & 0.000049 & 5329.570282 & 0.000029 \\
4558.521225 & 0.000036 & 4936.575635 & 0.000023 & 5246.582689 & 0.000029 & 5330.504008 & 0.000027 \\
\hline
\end{tabular}
\end{table*}

\begin{table*}
\caption{Additional SuperWASP times of minimum light for HW~Vir (excluded from analysis here).}
\centering
\begin{tabular}{l l | l l | l l | l l}
\hline\hline
Min. time & Uncertainty & Min. time & Uncertainty & Min. time & Uncertainty & Min. time & Uncertainty \\
(BJD$-$2450000) & (d) & (BJD$-$2450000) & (d) & (BJD$-$2450000) & (d) & (BJD$-$2450000) & (d) \\
\hline
4305.468229 & 0.000096 & 4587.583901 & 0.000206 & 4886.502134 & 0.000121 & 5252.535517 & 0.000083 \\
4307.574526 & 0.000034 & 4591.552963 & 0.000021 & 4889.537571 & 0.000095 & 5253.468562 & 0.000100 \\
4502.495970 & 0.000071 & 4592.487433 & 0.000496 & 4896.540716 & 0.000090 & 5265.491302 & 0.000091 \\
4515.568553 & 0.000068 & 4593.537082 & 0.000096 & 4908.562921 & 0.000124 & 5272.494266 & 0.000228 \\
4516.505738 & 0.000162 & 4596.571862 & 0.000168 & 4909.495996 & 0.000086 & 5289.537014 & 0.000159 \\
4525.489930 & 0.000039 & 4598.556117 & 0.000076 & 4916.500054 & 0.000096 & 5290.469244 & 0.000094 \\
\hline
\end{tabular}
\end{table*}

\begin{table*}
\caption{Selected good SuperWASP times of minimum light for 2M~1938+4603.}
\centering
\begin{tabular}{l l | l l | l l | l l}
\hline\hline
Min. time & Uncertainty & Min. time & Uncertainty & Min. time & Uncertainty & Min. time & Uncertainty \\
(BJD$-$2450000) & (d) & (BJD$-$2450000) & (d) & (BJD$-$2450000) & (d) & (BJD$-$2450000) & (d) \\
\hline
3128.789803 & 0.000895 & 4272.623975 & 0.000525 & 4609.549984 & 0.001188 & 4661.616154 & 0.000849 \\
3129.543451 & 0.001087 & 4274.509563 & 0.000853 & 4613.573371 & 0.000661 & 4662.622276 & 0.000891 \\
3130.548944 & 0.001572 & 4275.516890 & 0.000867 & 4614.579912 & 0.000526 & 4663.502816 & 0.000400 \\
3132.562759 & 0.001656 & 4276.523378 & 0.000853 & 4629.544976 & 0.000508 & 4664.507549 & 0.000775 \\
3137.592372 & 0.000472 & 4277.528190 & 0.000579 & 4630.551936 & 0.000509 & 4665.514457 & 0.001457 \\
3138.598494 & 0.000916 & 4278.535098 & 0.001781 & 4631.559944 & 0.001324 & 4666.521312 & 0.001301 \\
\hline
\end{tabular}
\end{table*}

\begin{table*}
\caption{Additional SuperWASP times of minimum light for 2M~1938+4603 (excluded from analysis here).}
\centering
\begin{tabular}{l l | l l | l l | l l}
\hline\hline
Min. time & Uncertainty & Min. time & Uncertainty & Min. time & Uncertainty & Min. time & Uncertainty \\
(BJD$-$2450000) & (d) & (BJD$-$2450000) & (d) & (BJD$-$2450000) & (d) & (BJD$-$2450000) & (d) \\
\hline
4234.525448 & 0.002080 & 4293.496128 & 0.003736 & 4651.554518 & 0.001352 & 5349.552322 & 0.001912 \\
4235.519517 & 0.001039 & 4294.512363 & 0.000712 & 4652.561687 & 0.000685 & 5360.626159 & 0.001240 \\
4254.503881 & 0.001853 & 4620.623507 & 0.001283 & 4667.528692 & 0.000575 & 5382.496992 & 0.001102 \\
4269.609384 & 0.000545 & 4621.624649 & 0.000736 & 4672.559981 & 0.000528 &  &  \\
4270.609270 & 0.000556 & 4622.504115 & 0.001623 & 4675.576827 & 0.000550 &  &  \\
4273.499930 & 0.000697 & 4649.549611 & 0.000545 & 4687.523579 & 0.000784 &  &  \\
\hline
\end{tabular}
\end{table*}

\begin{table*}
\caption{Selected good SuperWASP times of minimum light for NSVS~07826147.}
\centering
\begin{tabular}{l l | l l | l l | l l}
\hline\hline
Min. time & Uncertainty & Min. time & Uncertainty & Min. time & Uncertainty & Min. time & Uncertainty \\
(BJD$-$2450000) & (d) & (BJD$-$2450000) & (d) & (BJD$-$2450000) & (d) & (BJD$-$2450000) & (d) \\
\hline
3130.529723 & 0.000247 & 3223.547864 & 0.000209 & 4228.627436 & 0.000117 & 5676.634771 & 0.000098 \\
3132.632638 & 0.000231 & 3225.488789 & 0.000219 & 4230.568681 & 0.000193 & 5677.605048 & 0.000280 \\
3135.544759 & 0.000279 & 3226.621132 & 0.000068 & 4231.539506 & 0.000250 & 5678.575974 & 0.000140 \\
3137.647791 & 0.000126 & 3227.591755 & 0.000069 & 4232.509960 & 0.000094 & 5684.561515 & 0.000235 \\
3138.618296 & 0.000088 & 3232.606773 & 0.000155 & 4233.642320 & 0.000151 & 5685.532171 & 0.000126 \\
3139.588767 & 0.000232 & 3233.577194 & 0.000103 & 4234.612968 & 0.000079 & 5686.502760 & 0.000145 \\
\hline
\end{tabular}
\end{table*}

\begin{table*}
\caption{Additional SuperWASP times of minimum light for NSVS~07826147 (excluded from analysis here).}
\centering
\begin{tabular}{l l | l l | l l | l l}
\hline\hline
Min. time & Uncertainty & Min. time & Uncertainty & Min. time & Uncertainty & Min. time & Uncertainty \\
(BJD$-$2450000) & (d) & (BJD$-$2450000) & (d) & (BJD$-$2450000) & (d) & (BJD$-$2450000) & (d) \\
\hline
3128.751225 & 0.000851 & 3242.635664 & 0.000212 & 3852.507483 & 0.000190 & 4206.634272 & 0.000402 \\
3129.559707 & 0.000252 & 3248.625451 & 0.000405 & 3854.610229 & 0.000078 & 4220.537835 & 0.001464 \\
3131.494819 & 0.000472 & 3249.592054 & 0.000110 & 3855.582790 & 0.000430 & 4221.508862 & 0.000522 \\
3140.559525 & 0.000144 & 3252.505751 & 0.000306 & 3856.551188 & 0.000353 & 4247.554713 & 0.000279 \\
3170.486383 & 0.000184 & 3253.637470 & 0.000279 & 3901.525833 & 0.000832 & 4248.525740 & 0.001133 \\
3175.501991 & 0.000122 & 3254.607419 & 0.000479 & 3903.630408 & 0.000083 & 4249.496228 & 0.000373 \\
\hline
\end{tabular}
\end{table*}

\begin{table*}
\caption{Selected good SuperWASP times of minimum light for NSVS~14256825.}
\centering
\begin{tabular}{l l | l l | l l | l l}
\hline\hline
Min. time & Uncertainty & Min. time & Uncertainty & Min. time & Uncertainty & Min. time & Uncertainty \\
(BJD$-$2450000) & (d) & (BJD$-$2450000) & (d) & (BJD$-$2450000) & (d) & (BJD$-$2450000) & (d) \\
\hline
4297.497367 & 0.000219 & 4354.450333 & 0.000223 & 5396.492988 & 0.000167 & 5454.549284 & 0.000268 \\
4298.491079 & 0.000244 & 4355.554075 & 0.000290 & 5397.486194 & 0.000228 & 5463.489591 & 0.000335 \\
4303.457962 & 0.000138 & 4356.547741 & 0.000119 & 5398.479538 & 0.000246 & 5691.523045 & 0.000285 \\
4304.451099 & 0.000142 & 4358.534579 & 0.000099 & 5399.472998 & 0.000331 & 5692.516206 & 0.000187 \\
4305.554588 & 0.000297 & 4359.527463 & 0.000176 & 5400.466319 & 0.000385 & 5693.509895 & 0.000208 \\
4306.548299 & 0.000216 & 4360.521176 & 0.000121 & 5408.523367 & 0.000153 & 5694.503377 & 0.000287 \\
\hline
\end{tabular}
\end{table*}

\begin{table*}
\caption{Additional SuperWASP times of minimum light for NSVS~14256825 (excluded from analysis here).}
\centering
\begin{tabular}{l l | l l | l l | l l}
\hline\hline
Min. time & Uncertainty & Min. time & Uncertainty & Min. time & Uncertainty & Min. time & Uncertainty \\
(BJD$-$2450000) & (d) & (BJD$-$2450000) & (d) & (BJD$-$2450000) & (d) & (BJD$-$2450000) & (d) \\
\hline
3904.789797 & 0.000201 & 4332.485465 & 0.000308 & 5055.547334 & 0.000059 & 5697.481503 & 0.000113 \\
3905.563566 & 0.000458 & 4333.479292 & 0.000154 & 5314.484489 & 0.000396 & 5698.476664 & 0.000531 \\
4302.463721 & 0.000158 & 4339.549916 & 0.000613 & 5332.474464 & 0.001272 & 5699.467502 & 0.000610 \\
4308.534873 & 0.000605 & 4340.543835 & 0.000426 & 5406.536621 & 0.001255 & 5713.488146 & 0.000515 \\
4310.521056 & 0.000623 & 4341.537203 & 0.001000 & 5407.527229 & 0.000684 & 5716.459326 & 0.000126 \\
4312.509078 & 0.000924 & 4342.531122 & 0.000344 & 5414.488298 & 0.000572 & 5724.527308 & 0.000242 \\
4313.499226 & 0.000754 & 4364.494610 & 0.000403 & 5418.457674 & 0.000259 & 5726.511559 & 0.000517 \\
4314.495261 & 0.000476 & 4368.462791 & 0.000364 & 5434.461925 & 0.000523 & 5753.554652 & 0.000444 \\
4315.488904 & 0.000307 & 4932.478734 & 0.000111 & 5439.538078 & 0.000275 & 5754.545122 & 0.000268 \\
4317.475547 & 0.000682 & 4934.460502 & 0.000385 & 5455.542836 & 0.000220 & 5755.539547 & 0.000627 \\
4318.468914 & 0.000459 & 4935.461319 & 0.000467 & 5462.496500 & 0.000532 & 5762.494452 & 0.000394 \\
4319.461499 & 0.000178 & 5045.504252 & 0.001249 & 5466.472797 & 0.000043 & 5770.550938 & 0.001184 \\
\hline
\end{tabular}
\end{table*}

\begin{table*}
\caption{Selected good SuperWASP times of minimum light for NY~Vir.}
\centering
\begin{tabular}{l l | l l | l l | l l}
\hline\hline
Min. time & Uncertainty & Min. time & Uncertainty & Min. time & Uncertainty & Min. time & Uncertainty \\
(BJD$-$2450000) & (d) & (BJD$-$2450000) & (d) & (BJD$-$2450000) & (d) & (BJD$-$2450000) & (d) \\
\hline
4515.530946 & 0.000190 & 4912.523711 & 0.000098 & 4992.528139 & 0.000242 & 5329.517385 & 0.000097 \\
4524.521364 & 0.000146 & 4913.533860 & 0.000093 & 4995.558785 & 0.000170 & 5330.527945 & 0.000233 \\
4527.552136 & 0.000197 & 4914.543840 & 0.000101 & 5000.508565 & 0.000124 & 5331.537956 & 0.000155 \\
4530.582403 & 0.000207 & 4915.554042 & 0.000103 & 5001.518641 & 0.000115 & 5332.548032 & 0.000136 \\
4533.511823 & 0.000185 & 4917.574213 & 0.000224 & 5009.498941 & 0.000100 & 5333.558149 & 0.000190 \\
4534.521961 & 0.000124 & 4918.584562 & 0.000106 & 5010.509205 & 0.000140 & 5334.568392 & 0.000185 \\
\hline
\end{tabular}
\end{table*}

\begin{table*}
\caption{Additional SuperWASP times of minimum light for NY~Vir (excluded from analysis here).}
\centering
\begin{tabular}{l l | l l | l l | l l}
\hline\hline
Min. time & Uncertainty & Min. time & Uncertainty & Min. time & Uncertainty & Min. time & Uncertainty \\
(BJD$-$2450000) & (d) & (BJD$-$2450000) & (d) & (BJD$-$2450000) & (d) & (BJD$-$2450000) & (d) \\
\hline
4307.540009 & 0.000168 & 4601.596771 & 0.000362 & 4927.574474 & 0.000397 & 5283.555410 & 0.000410 \\
4516.540180 & 0.000489 & 4606.547225 & 0.000520 & 4929.596013 & 0.000579 & 5284.565570 & 0.000489 \\
4525.532196 & 0.000460 & 4607.556669 & 0.000508 & 4933.535171 & 0.000577 & 5288.505780 & 0.000228 \\
4532.502211 & 0.000261 & 4608.566702 & 0.000621 & 4934.544110 & 0.000771 & 5296.585708 & 0.000203 \\
4535.532942 & 0.000047 & 4609.576861 & 0.000869 & 4938.585274 & 0.000201 & 5309.516799 & 0.000393 \\
4543.506719 & 0.000892 & 4611.598358 & 0.000309 & 4943.536044 & 0.000209 & 5316.587872 & 0.000637 \\
\hline
\end{tabular}
\end{table*}

\clearpage

\begin{table*}
\caption{Selected good SuperWASP times of minimum light for DE~CVn.}
\centering
\begin{tabular}{l l | l l | l l | l l}
\hline\hline
Min. time & Uncertainty & Min. time & Uncertainty & Min. time & Uncertainty & Min. time & Uncertainty \\
(BJD$-$2450000) & (d) & (BJD$-$2450000) & (d) & (BJD$-$2450000) & (d) & (BJD$-$2450000) & (d) \\
\hline
3130.484921 & 0.003768 & 3183.651996 & 0.001788 & 4161.364177 & 0.001634 & 4223.632158 & 0.002288 \\
3131.575822 & 0.002578 & 3185.472313 & 0.001355 & 4163.547951 & 0.002535 & 4224.725183 & 0.003060 \\
3132.670895 & 0.002256 & 3190.570188 & 0.002987 & 4165.371076 & 0.003973 & 4225.452551 & 0.002077 \\
3137.404479 & 0.001359 & 3192.391036 & 0.001608 & 4166.457425 & 0.002259 & 4227.637842 & 0.001696 \\
3138.498414 & 0.001410 & 3193.482392 & 0.001660 & 4167.555002 & 0.002034 & 4228.365969 & 0.002451 \\
3139.589922 & 0.001497 & 3195.671249 & 0.001512 & 4171.560724 & 0.000761 & 4230.551109 & 0.001067 \\
\hline
\end{tabular}
\end{table*}

\begin{table*}
\caption{Additional SuperWASP times of minimum light for DE~CVn (excluded from analysis here).}
\centering
\begin{tabular}{l l | l l | l l | l l}
\hline\hline
Min. time & Uncertainty & Min. time & Uncertainty & Min. time & Uncertainty & Min. time & Uncertainty \\
(BJD$-$2450000) & (d) & (BJD$-$2450000) & (d) & (BJD$-$2450000) & (d) & (BJD$-$2450000) & (d) \\
\hline
3128.674238 & 0.005462 & 3184.388771 & 0.003227 & 3229.540683 & 0.000784 & 4192.691541 & 0.003178 \\
3135.563451 & 0.001377 & 3188.373669 & 0.001225 & 3231.724457 & 0.000997 & 4203.603887 & 0.003905 \\
3140.654726 & 0.002351 & 3189.485508 & 0.003859 & 3233.557405 & 0.000626 & 4204.711478 & 0.002353 \\
3155.611142 & 0.005107 & 3191.663061 & 0.001341 & 3240.445594 & 0.001316 & 4205.414267 & 0.006544 \\
3156.717822 & 0.004115 & 3194.562217 & 0.001511 & 3242.653189 & 0.000154 & 4211.615560 & 0.005802 \\
3160.712431 & 0.012446 & 3198.571998 & 0.002952 & 4135.511042 & 0.051272 & 4216.728683 & 0.004243 \\
3167.643065 & 0.002421 & 3199.663203 & 0.001589 & 4139.527499 & 0.001468 & 4220.738009 & 0.001626 \\
3168.737152 & 0.003400 & 3202.586938 & 0.001087 & 4140.617793 & 0.001723 & 4226.548307 & 0.006701 \\
3172.372173 & 0.001837 & 3206.598843 & 0.000667 & 4146.425209 & 0.002043 & 4235.649667 & 0.004393 \\
3175.633344 & 0.001819 & 3207.701123 & 0.000779 & 4150.451983 & 0.003303 & 4247.668389 & 0.003734 \\
3176.348271 & 0.001484 & 3209.482220 & 0.003281 & 4154.459336 & 0.002896 & 4251.695771 & 0.004326 \\
3179.650104 & 0.073804 & 3217.536679 & 0.001151 & 4158.434524 & 0.002608 & 4258.589536 & 0.000588 \\
3180.358962 & 0.002195 & 3221.533108 & 0.000607 & 4169.372361 & 0.004738 & 4263.710549 & 0.001383 \\
3182.550474 & 0.002398 & 3225.524909 & 0.000232 & 4170.484654 & 0.001754 &  &  \\
\hline
\end{tabular}
\end{table*}

\begin{table*}
\caption{Selected good SuperWASP times of minimum light for QS~Vir.}
\centering
\begin{tabular}{l l | l l | l l | l l}
\hline\hline
Min. time & Uncertainty & Min. time & Uncertainty & Min. time & Uncertainty & Min. time & Uncertainty \\
(BJD$-$2450000) & (d) & (BJD$-$2450000) & (d) & (BJD$-$2450000) & (d) & (BJD$-$2450000) & (d) \\
\hline
4623.455406 & 0.000421 & 4937.482980 & 0.000467 & 5010.600789 & 0.000311 & 5324.477971 & 0.000509 \\
4624.510960 & 0.000409 & 4939.593901 & 0.000324 & 5011.505021 & 0.000310 & 5329.603478 & 0.000176 \\
4627.526112 & 0.000221 & 4940.498258 & 0.000745 & 5022.511205 & 0.000245 & 5331.563515 & 0.000164 \\
4628.580598 & 0.000350 & 4941.553812 & 0.000419 & 5245.480741 & 0.000237 & 5332.467873 & 0.000244 \\
4641.547196 & 0.000517 & 4942.457808 & 0.000224 & 5246.536326 & 0.000266 & 5333.522736 & 0.000408 \\
4642.601117 & 0.000234 & 4943.513661 & 0.000528 & 5248.495201 & 0.000536 & 5334.579107 & 0.000378 \\
\hline
\end{tabular}
\end{table*}

\begin{table*}
\caption{Additional SuperWASP times of minimum light for QS~Vir (excluded from analysis here).}
\label{qsvirextra}
\centering
\begin{tabular}{l l | l l | l l | l l}
\hline\hline
Min. time & Uncertainty & Min. time & Uncertainty & Min. time & Uncertainty & Min. time & Uncertainty \\
(BJD$-$2450000) & (d) & (BJD$-$2450000) & (d) & (BJD$-$2450000) & (d) & (BJD$-$2450000) & (d) \\
\hline
4307.619507 & 0.000252 & 4938.536273 & 0.000774 & 5255.582127 & 0.000862 & 5352.520076 & 0.000199 \\
4626.462015 & 0.000729 & 4944.566577 & 0.001012 & 5256.485793 & 0.000600 & 5353.575474 & 0.000264 \\
4640.486616 & 0.001555 & 4946.529441 & 0.000921 & 5265.534893 & 0.000204 & 5366.538836 & 0.000829 \\
4649.537663 & 0.000264 & 4952.562885 & 0.003692 & 5272.466727 & 0.001010 & 5367.594642 & 0.000573 \\
4651.491293 & 0.000709 & 4954.524242 & 0.000838 & 5273.519140 & 0.000884 & 5372.569516 & 0.000829 \\
4662.487488 & 0.000439 & 4956.477809 & 0.000865 & 5283.471654 & 0.000731 & 5373.473811 & 0.000652 \\
\hline
\end{tabular}
\end{table*}

\end{appendix}

\end{document}